\documentclass{he_symp}
\usepackage{psfig,graphicx,epsfig}
\usepackage{color}
\usepackage{amsmath,amssymb,epic,eepic,array}
\begin{document}
\renewcommand{\FirstPageOfPaper }{ 167}\renewcommand{\LastPageOfPaper }{ 170}

\title{The Characteristics of (Normal) Pulsars}
\author{Richard Wielebinski}
\institute{Max--Planck--Institut f\"ur Radioastronomie,
 Auf dem H\"ugel 69, 53121 Bonn, Germany}
\maketitle

\begin{abstract}
Pulsars have now been studied for 34 years. We know of the existence
of some 1500 objects at radio frequencies. Many of the characteristics
of pulsars such as pulsar period, period derivative, spectrum,
polarization, etc., have been catalogued for many objects. Still we do
not know the details of the pulsar emission mechanism.  The present
review will give an up-to-date description of the characteristics of
(normal, slow) radio pulsars. Millisecond pulsars will be dealt with by
M.~Kramer (this volume).
\end{abstract}

\section{Introduction}
The discovery of pulsars by Hewish et al., made at the low radio
frequency of 81.5~MHz, stunned the astronomical community. In the
discovery paper the authors speculated already on the possibility that
we had either white dwarfs or neutron stars showing the pulsations.
Virtually every radio telescope around the world was observing pulsars
in 1968. Many of the important parameters of pulsars like the exact
period, the period derivative, pulse shape, the polarization and
spectrum were determined in the earliest papers already. Also the
Galactic distribution of pulsars was found after the discoveries with
the Molonglo telescope (Large et al. 1968a) increased rapidly the
number of known objects. The Pulsar -- Supernova correlation was
suggested by the discovery of PSR~0833$-$45 near the Vela~X SNR (Large
et al. 1968b) and conclusively confirmed by the discovery of the pulsar
in the Crab nebula (Clomella et al. 1969; Zeissig \& Richards 1969).
This gave the theory of pulsar formation great impulse. The
measurement of the pulsar polarization in PSR~0833$-$45 by Radhakrishnan
\& Cooke (1969), that showed a vector sweep, finally clinched the
argument that we were observing the magnetic poles of a rotating
neutron star. Further important discoveries at radio frequencies
were the binary pulsar (Hulse \& Taylor 1975), the millisecond pulsar
(Backer et al. 1982) and the planets around a pulsar (Wolszczan
\& Frail 1992).

The early discoveries made way to large search programmes and to
surveys of pulsar characteristics. All the large radio telescopes have
been involved in finding new objects. In particular, the recent Parkes
multibeam survey (e.g. Manchester et al. 2001), led to the discovery of
more than 600 new pulsars. To determine the pulsar spectra we require
the collection of data over a wide frequency range. In this area the
Effelsberg radio telescope was most prominent since it was the only
instrument able to observe pulsars at high radio frequencies. Pulsar
polarization is  a very important parameter that has to be known for
any serious emission theory discussion. This was observed by many
telescopes but also in particular in Effelsberg. Pulsar emission shows
many details, like drifting sub-pulse emission, pulsar modes or
microstructure. I will not discuss these details because of lack of
time. I will concentrate on pointing out what I consider to be the
important connections of known pulsar parameters. Towards the end of
this talk I will describe some promising new avenues of observational
pulsar research.

\section{Pulsar period}
The periods of pulsars are remarkably stable, in the range 0.0015~sec
to 8.51~sec. The pulsar period has been found to change (increase) as a
result of loss of rotational energy. This period derivative P(dot) is a
very important parameter for the determination of the pulsar age. The
`normal' pulsars are found in the period range of 0.033~sec for the
Crab pulsar and up to 8.51~sec for PSR~J2144$-$3933 (Young et al. 2000).
The peak of the period distribution of pulsars lies around 0.7~sec.
A second population of pulsars, the millisecond pulsars, is to be found
in the period range 1.558 millisec to 20 millisec. These objects have
very small period derivative and are considered to be `recycled'
pulsars. These are the best clocks in the universe.

\begin{figure*}
\begin{minipage}[b]{8cm}
\psfig{figure=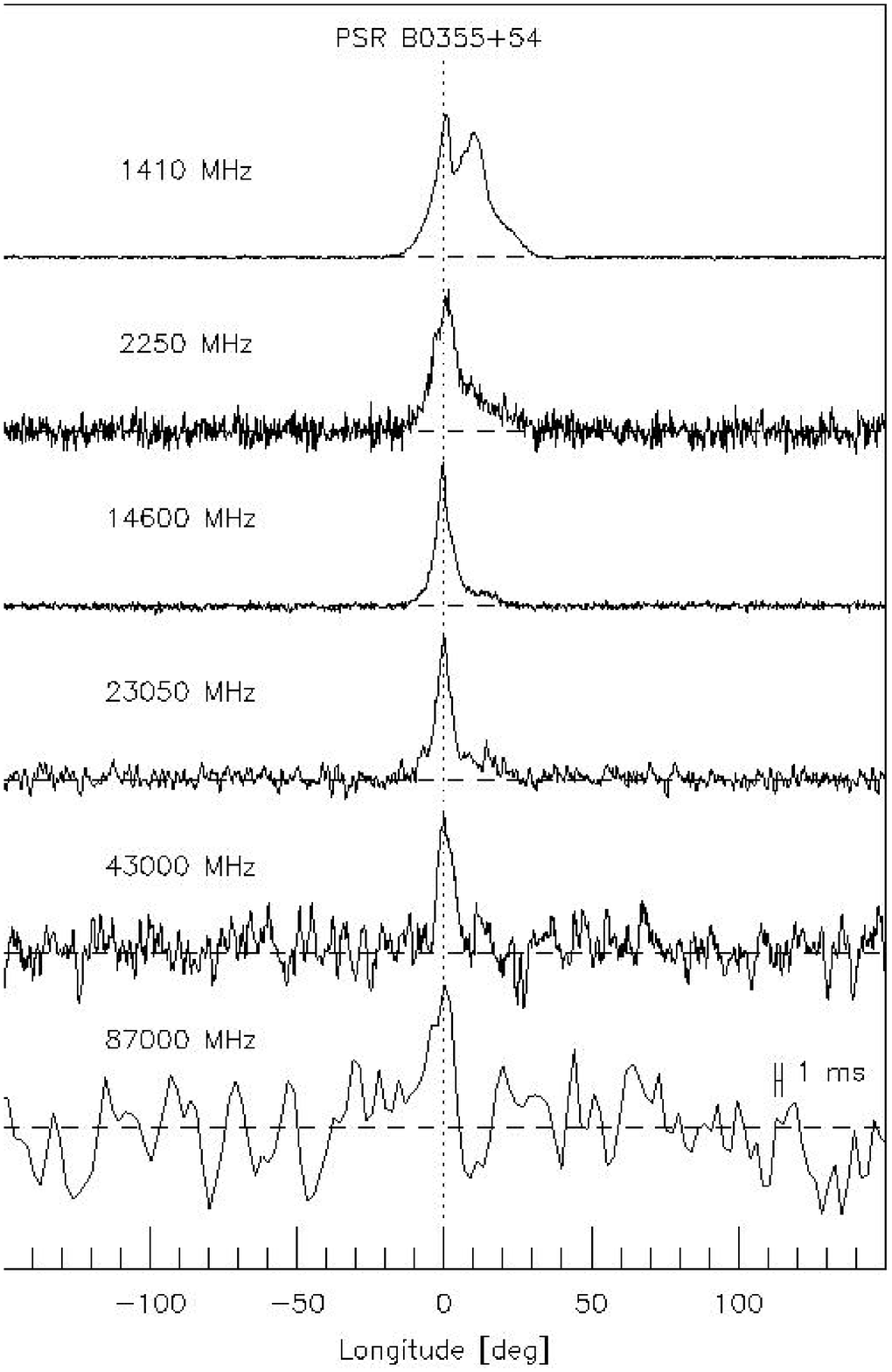,width=8cm,clip=}
\caption{Pulse shape of PSR B0355+54 aligned by timing}
\end{minipage}\hfill
\begin{minipage}[b]{8cm}
\psfig{figure=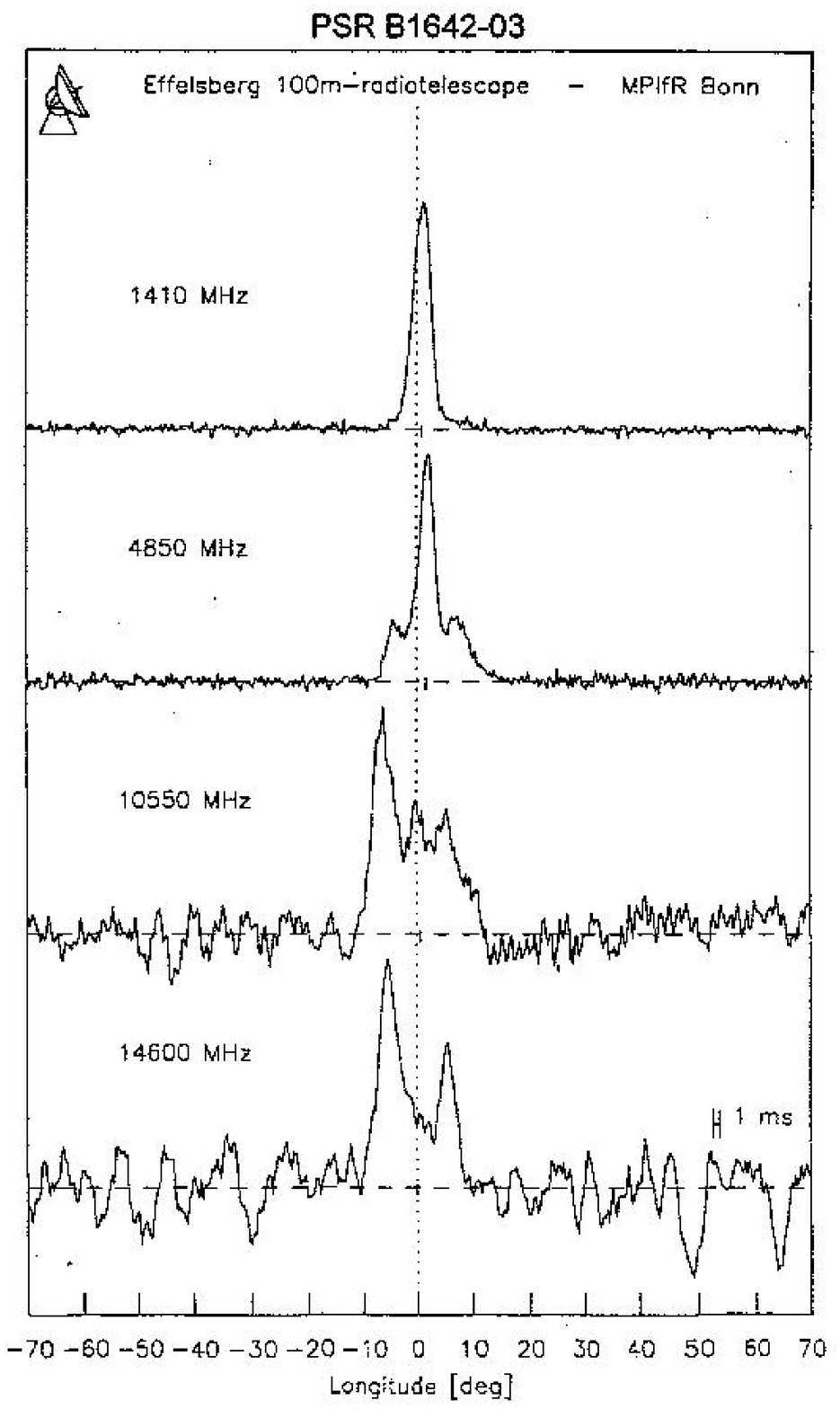,width=8truecm,clip=} \caption{Pulse shape
of PSR B1642$-$03. Note the unusual frequency evolution of this
pulsar}
\end{minipage}
\end{figure*}

\section{Pulse shapes}

One of the very characteristic parameters of a radio pulsar is its
pulse shape. The pulse shape is due to the emission region that is
close to the magnetic pole in a dipolar magnetic field of the rotating
neutron star. This emission region sweeps past the observer giving,
depending on the geometrical orientations of the rotation axis and the
magnetic axis forming a pulsar beam. It is common knowledge that after
the addition of some thousand of pulses a stable pulse shape appears.
This pulse to space ratio is mostly a few \% with some exceptions. Some
pulsars have a duty ratio of more than 50\%. The pulse shapes can be
single double, triple or even more complex. The single pulse shape is
considered to be due to an inner `core'. In triple pulse shapes we
observe the `core' and two `cones' being cut by the line of sight. The
most complex pulse shapes show up to nine components. Interpulses are
observed in some pulsars, being interpreted as the emission from the
opposite pole. The stability of pulse shapes has to be taken with a
great BUT! When an observer can discern single pulses a huge variety of
behavior is seen. Each single pulse is formed from a series of
sub-pulses. These sub-pulses can show microstructure which seems to be
limited only by the temporal resolution of the recording equipment.
Sub-pulses can drift within the integrated pulse shape envelope giving
multiple (non-integral) periodicities. The description of all these
effects becomes even more complicated when one considers the evolution
of pulse shape with radio frequency. The general evolution is to a
narrower pulse shape at higher radio frequencies that is often cited as
the Radius--to--Frequency mapping. At higher radio frequencies emission
is thought to come from regions closer to the neutron star surface. A
large collection of pulse shapes is to be found in Seiradakis et al.
(1995) and in Kijak et al. (1998). The evolution of pulse shape with
frequency is documented in Sieber et al. (1975), Kramer et al. (1997)
or Kuzmin et al. (1998). I present the pulse shapes for the pulsars
0355+54 (Fig.~1) and 1642$-$03 (Fig.~2) to illustrate the different
evolutions of pulse shape with frequency. Finally there are moding
pulsars where a stable pulse shape changes for some time to an abnormal
mode, returning to normal after an unpredictable length of time.

\section{Pulsar spectra}

Observing at five frequencies (four of them simultaneously) with the
Parkes telescope, Robinson et al. (1968) established the spectrum of
CP1919+21 from 85~MHz to 2.7~GHz. This `classical' spectrum showed a
steep spectral index at decimeter wavelengths ($\alpha \sim 1.5$)
with a spectral break  with the spectral index increasing to $\alpha
\sim 3.0$ at the highest frequencies. The continuation of gathering of
data at many frequencies was a prerequisite of determining pulsar
spectra. Here a serious problem must be mentioned, namely that there
are huge flux variations due to interstellar scattering, but possibly
also due to internal variations in the pulsar emission. Pulsars with
high dispersion measure exhibit particularly strong `fading' at
high radio frequencies. To determine a flux of a pulsar at a frequency
repeated observations are necessary. In spite of these difficulties
several large data sets are available (e.g. Malofeev et al. 1994;
Lorimer et al. 1995; Toscano et al. 1998; Maron et al. 2000). The
older data on pulsar spectra were derived for flux density measurements
in the frequency range 85~MHz to 1400~MHz. The newer data have extended
the spectra of hundreds of pulsars up to 4.8~GHz and down to 102~MHz.
For the strongest objects data are available at 10~GHz and even at mm
wavelengths. Here a surprise has been found: the spectrum of some
pulsars did not continue fall but showed a surprising `flattening' or
even `turn-up' (e.g. Wielebinski et al. 1993; Kramer et al. 1997). The
spectra of eight strongest pulsars with detected flux values at mm
wavelengths are shown in Figure~3. One pulsar could be detected at
87~GHz (Morris et al. 1997). A change in the pulsar emission mechanism?

\section{Pulsar polarization}

The earliest observations (e.g. Lyne \& Smith 1968) showed that
sub-pulse components had very high degree of polarization. At first it
was only the linear polarization that was studied showing the
characteristic position angle sweep that indicated rotating objects.
Later circular polarization was also detected (e.g. Manchester 1971). A
surprising result was the observation by Manchester et al. (1973) that
pulsar polarization remains high up to some critical frequency and then
falls rapidly. Orthogonal polarization modes were postulated (e.g.
Manchester 1975) based on the observation that in some pulsars the
position angle sweep has jumps of $\sim 90\degr$. Recent surveys of
integrated polarization of pulsars at many frequencies (e.g. von
Hoensbroech \& Xilouris 1997; Gould \& Lyne 1998; Weisberg et al. 1999;
von Hoensbroech 1999) showed us the great diversity in polarization
morphology. A type of pulsar has been discovered (von Hoensbroech
1999) in which the linear polarization falls to high frequencies with a
corresponding increase of circular polarization. This was interpreted
as a propagation effect (von Hoensbroech \& Lesch 1999). The early
conclusion that the linear polarization drops suddenly after reaching
some critical frequency has been confirmed. This is shown in Figure~4.
This drop in linear polarization seem to occur when the spectrum
(compare Figures 3 and 4) of the pulsar changes. Does the loss of
coherence (polarization) occur when another emission mechanism
(incoherent one) begins to dominate?

\begin{figure*}
\begin{minipage}[b]{8.4cm}
\psfig{file=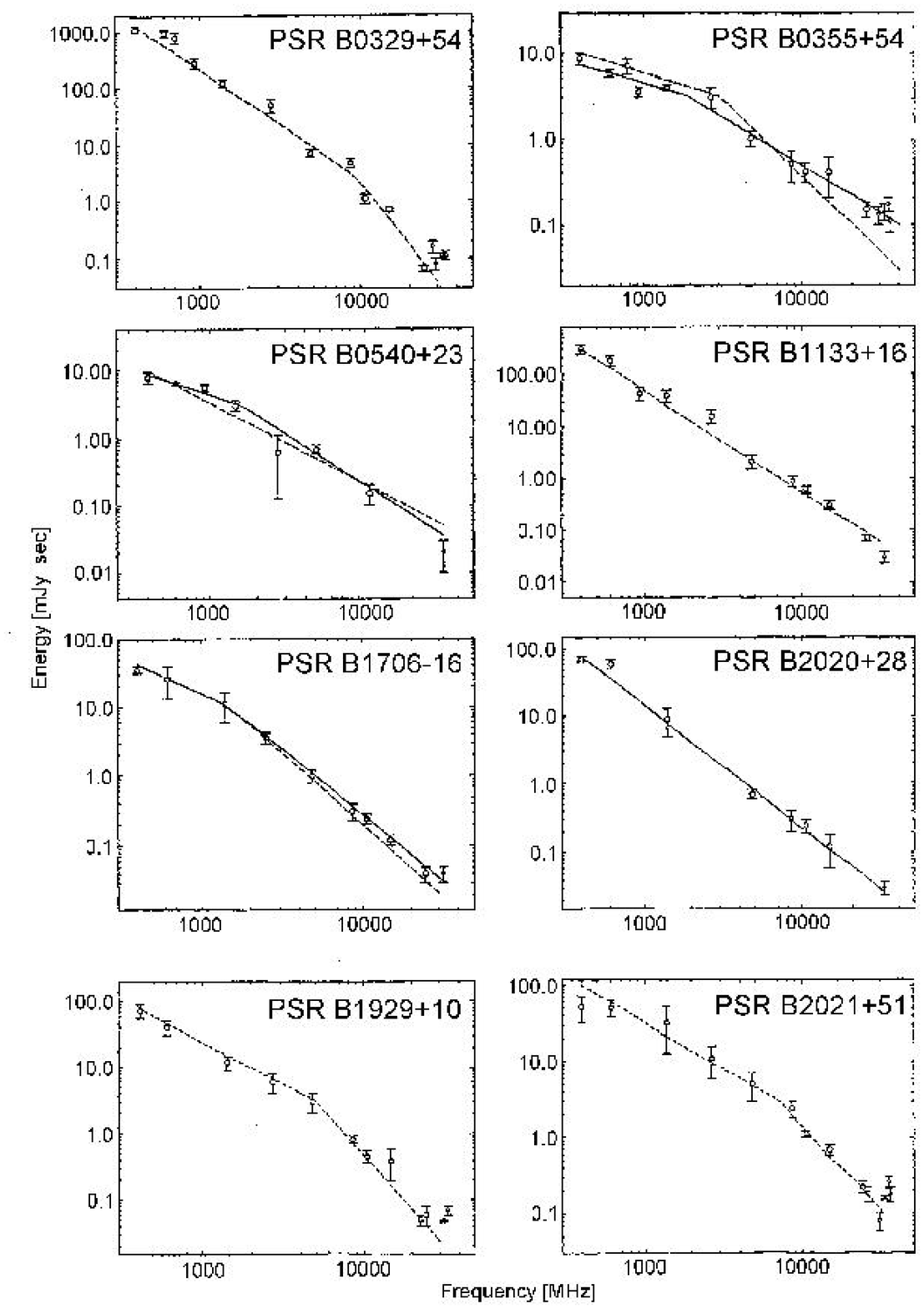,width=8cm,clip=}
\caption{The spectra of eight pulsars detected at mm-wavelengths}
\end{minipage}\hfill
\begin{minipage}[b]{8cm}
\psfig{file=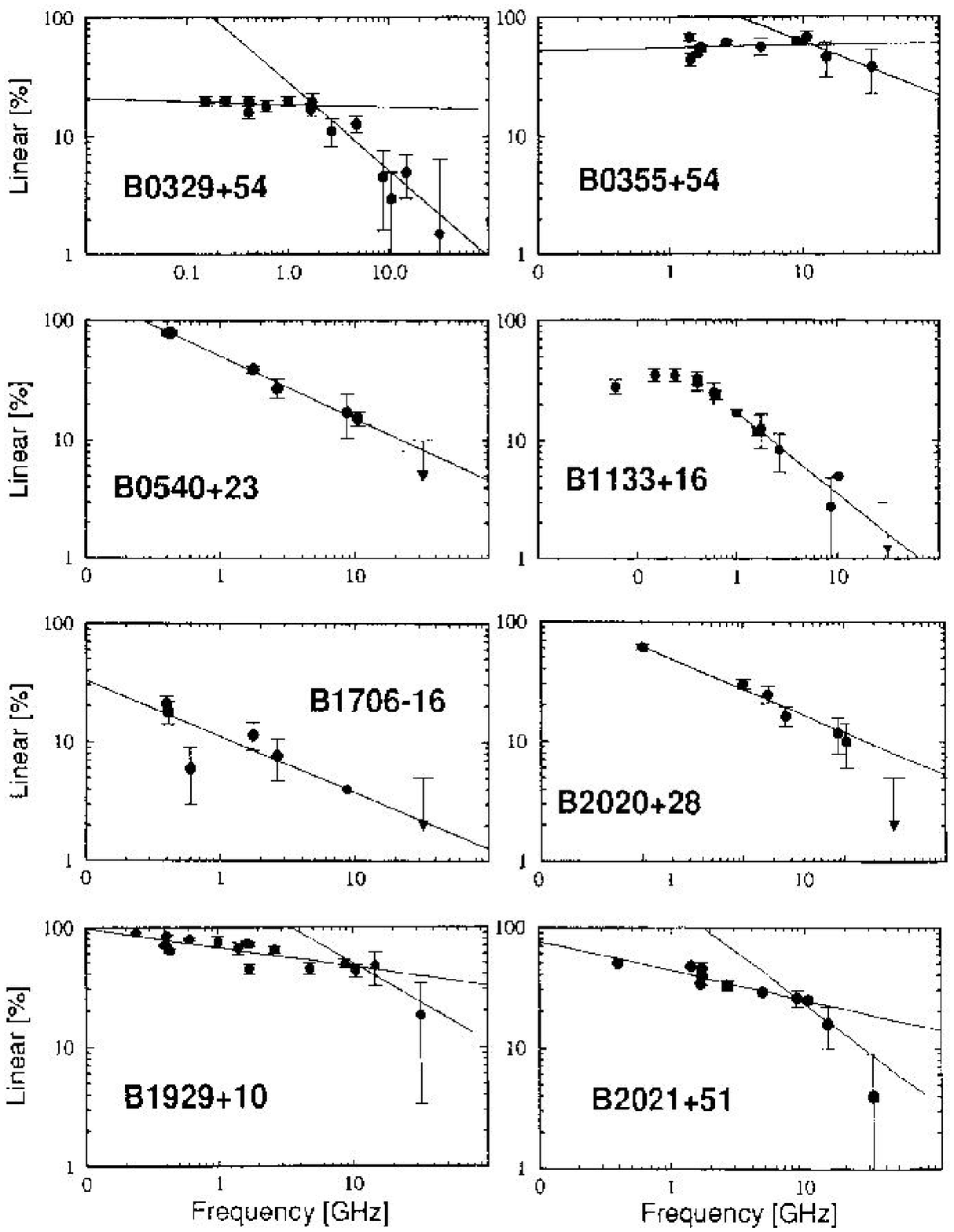,width=8.4cm,clip=}
\caption{The polarization of the eight pulsars shown in Figure 3. Note
the sudden break in polarization and the corresponding change in the
spectrum}
\end{minipage}
\end{figure*}

\section{Discussion}

In this talk many of the integral characteristics of pulsars were
discussed. I presented a summary about pulsar periods, their spectra,
the polarization and pulse shapes. I have not dealt with the
characteristics of millisecond pulsars since they are presented
elsewhere in this volume. I want to bring to your attention the
importance of studying single pulses at many frequencies
simultaneously. Such observations require the use of many telescopes
simultaneously. A contribution by Karastergiou (this volume) describes
recent four-station experiments. The single pulse data, with full
polarization, taken over a wide frequency range may hold the key to
understanding the pulsar emission process. Also the extension of pulsar
observations to yet higher radio frequencies (into the sub-mm
wavelength range) may finally allow us to `join' the radio to infrared
spectra and hence understand how pulsars work.

\clearpage

\end{document}